\documentclass[lettersize,journal]{IEEEtran}
\usepackage{amsmath,amsfonts}
\usepackage{algorithmic}
\usepackage{algorithm}
\usepackage{array}
\usepackage[caption=false,font=normalsize,labelfont=sf,textfont=sf]{subfig}
\usepackage{textcomp}
\usepackage{stfloats}
\usepackage{url}
\usepackage{verbatim}
\usepackage{graphicx}
\usepackage{cite}
\usepackage{tabularx}
\usepackage{booktabs}
\usepackage{tikz}

\definecolor{pipelineblue}{RGB}{16,54,111}

\newcommand{\stagenum}[1]{%
    \tikz[baseline=(n.base)]{
        \node[
            circle,
            fill=pipelineblue,
            text=white,
            font=\bfseries,
            minimum size=1.4em,
            inner sep=0.5pt
        ] (n) {#1};
    }%
}

\begin{document}

\title{Real-Time 250-kHz Spectral-Domain Optical Coherence Tomography Using a Jetson AGX Orin}

\author{Yash Nakadi, Austin Yetter, Toast Lobel, and Jungeun Won%
\thanks{Y. Nakadi is with the Department of Computer Science and Engineering, University at Buffalo, Buffalo, NY 14260 USA (e-mail: ynakadi@buffalo.edu).}%
\thanks{A. Yetter is with the Department of Biomedical Engineering, University at Buffalo, Buffalo, NY 14260 USA (e-mail: austinye@buffalo.edu).}%
\thanks{T. Lobel is with the Department of Biomedical Engineering, University at Buffalo, Buffalo, NY 14260 USA (e-mail: ajlobel@buffalo.edu).}%
\thanks{J. Won is with the Department of Biomedical Engineering, University at Buffalo, Buffalo, NY 14260 USA (e-mail: jungeunw@buffalo.edu).}}

\markboth{IEEE Transactions on Instrumentation and Measurement}%
{Nakadi \MakeLowercase{\textit{et al.}}: Embedded 250-kHz Spectral-Domain Optical Coherence Tomography}


\maketitle

\begin{abstract}
High-performance optical coherence tomography (OCT) relies on desktop workstations for data acquisition, reconstruction, and visualization, limiting integration into compact imaging systems. We evaluated whether an embedded GPU platform could sustain real-time operation of a 250-kHz spectral-domain OCT instrument without a desktop computer. A custom C++/CUDA OCT engine running on a NVIDIA Jetson AGX Orin 64 GB was developed and integrated with a 2048-pixel spectrometer, Camera Link frame grabber, and galvanometer scanner, and compared against an RTX A4500 workstation under the same scan protocols. Both systems maintained mean B-scan processing latency well below the corresponding B-scan acquisition intervals. Although the RTX A4500 workstation achieved lower reconstruction latency, Jetson eliminated explicit host-to-device transfer and exhibited more consistent processing latency. Furthermore, the measured active wall power was 67 W for the Jetson and 155 W for the workstation, corresponding to approximately 2.3 times greater wall-energy efficiency for the Jetson. In a separate 62.6-min GUI-enabled measurement, the Jetson processed 800,000 B-scans without frame loss while maintaining real-time reconstruction and visualization. These results demonstrate that embedded computing can support high-speed OCT acquisition, processing, and control while reducing wall-power demand and computing-platform size.
\end{abstract}

\begin{IEEEkeywords}
Optical coherence tomography, embedded GPU computing, real-time processing, GPU acceleration, NVIDIA Jetson, energy efficiency.  
\end{IEEEkeywords}

\section{Introduction}
\IEEEPARstart{O}{ptical} coherence tomography (OCT) is a non-invasive optical imaging technique that generates depth-resolved visualization of samples at micrometer-scale resolution using low coherence interferometry \cite{RN1}. OCT has become an indispensable tool for the diagnosis and management of retinal diseases \cite{RN33}. Modern commercial ophthalmic OCT systems are increasingly operating in 100-250 kHz A-scan rates, with newer devices that extend to 400-kHz, to support larger fields of view, faster volumetric imaging, and repeated acquisition of B-scans for OCT angiography (OCTA) \cite{RN20,RN24,RN25}. Research prototypes have further pushed imaging speeds into the several-hundred-kHz to the MHz A-scan rate regime \cite{RN12,RN22,RN23}. Together, these developments have increased the computational demands associated with the acquisition, reconstruction, and visualization of OCT data. GPU acceleration has enabled real-time image reconstruction and visualization in high-performance systems \cite{RN19,RN40}. However, these capabilities are often achieved using desktop or workstation-class computers equipped with discrete CPUs and GPUs.

At the same time, reducing the size, cost, and power consumption of OCT instrumentation is important for expanding imaging access beyond retinal specialty practices and research settings. Commercial systems from Lumedica and Philophos also demonstrate the practical relevance of compact and affordable OCT technologies \cite{RN11,RN15}. Portable systems, including handheld probes and endoscopes, low-cost platforms, and stand-alone imaging systems, have been developed for point-of-care, bedside, and resource-limited settings \cite{RN7,RN9,RN11,RN26,RN36}. While device miniaturization and portability can contribute to lower costs, portability and affordability represent distinct design objectives. Many handheld OCT probes utilize external workstation-class computers to achieve high-performance imaging, requiring cart-based systems. Although OCT systems based on Raspberry Pi \cite{RN34} and LattePanda \cite{RN27} single-board computers, mini-PC \cite{RN9}, and laptop \cite{RN8} have demonstrated compact and low-cost implementations, they may not satisfy the performance requirements of high-performance OCT systems.

Embedded computing platforms that integrate CPUs and GPUs within a compact system-on-module architecture may overcome this limitation. Recently, Wang et al. demonstrated a compact retinal OCT system based on the NVIDIA Jetson Orin Nano and reported OCT processing throughput exceeding 500-kHz while operating at an A-scan acquisition rate of 80-kHz \cite{RN15}. However, it remains unclear whether embedded platforms can scale to the acquisition rates and processing requirements of higher performance OCT systems.

In this work, we present and experimentally evaluate a 250-kHz SD-OCT instrument based on the NVIDIA Jetson AGX Orin. The system integrates high-bandwidth Camera Link acquisition through a PCIe frame grabber, camera-synchronized galvanometer control, and CUDA-based reconstruction with direct access to acquisition buffers. Unlike previous compact implementations operating at lower acquisition rates, the proposed architecture supports continuous acquisition and reconstruction at 250-kHz while providing local and network-accessible visualization. We characterize GPU execution time, data acquisition preparation, B-scan processing latency, wall-power demand, energy efficiency, and sustained operation. A matched visualization disabled implementation of the same reconstruction pipeline on a workstation provides a reference for evaluating the effects of GPU architecture, acquisition-memory organization, and system-level power consumption.

\section{Methods}
\subsection{Embedded SD-OCT system}
The OCT system was controlled by an NVIDIA Jetson AGX Orin 64 GB embedded computing platform (NVIDIA, Santa Clara, CA, USA). The Jetson incorporates a 12-core Arm Cortex-A78AE CPU, an NVIDIA Ampere GPU with 2048 CUDA cores, and 64 GB of shared LPDDR5 system memory providing 204.8 GB/s of memory bandwidth. Raw spectral data were acquired using a Cobra-S spectrometer (Wasatch Photonics, Morrisville, NC, USA) incorporating a 2048-pixel Teledyne e2v OctoPlus line-scan camera. The camera was operated in eight-tap, 10-bit mode with a 3-µs integration time at an A-scan rate of 250-kHz, using a Camera Link interface. This configuration produced a nominal 5.12-Gbit/s detector payload (640 MB/s), represented as a 1.024-GB/s 16-bit DMA stream in host memory. A BitFlow Axion 1xE frame grabber (BitFlow, Woburn, MA, USA) transferred the acquired data to the Jetson through a PCIe Gen2 ×4 interface. A superluminescent laser diode (Superlum Diodes Ltd., Carrigtwohill, Ireland) with a center wavelength of 850 nm and a full bandwidth of 144 nm was used as a light source.

\begin{figure}[!t]
    \centering
    \includegraphics[width=\columnwidth]{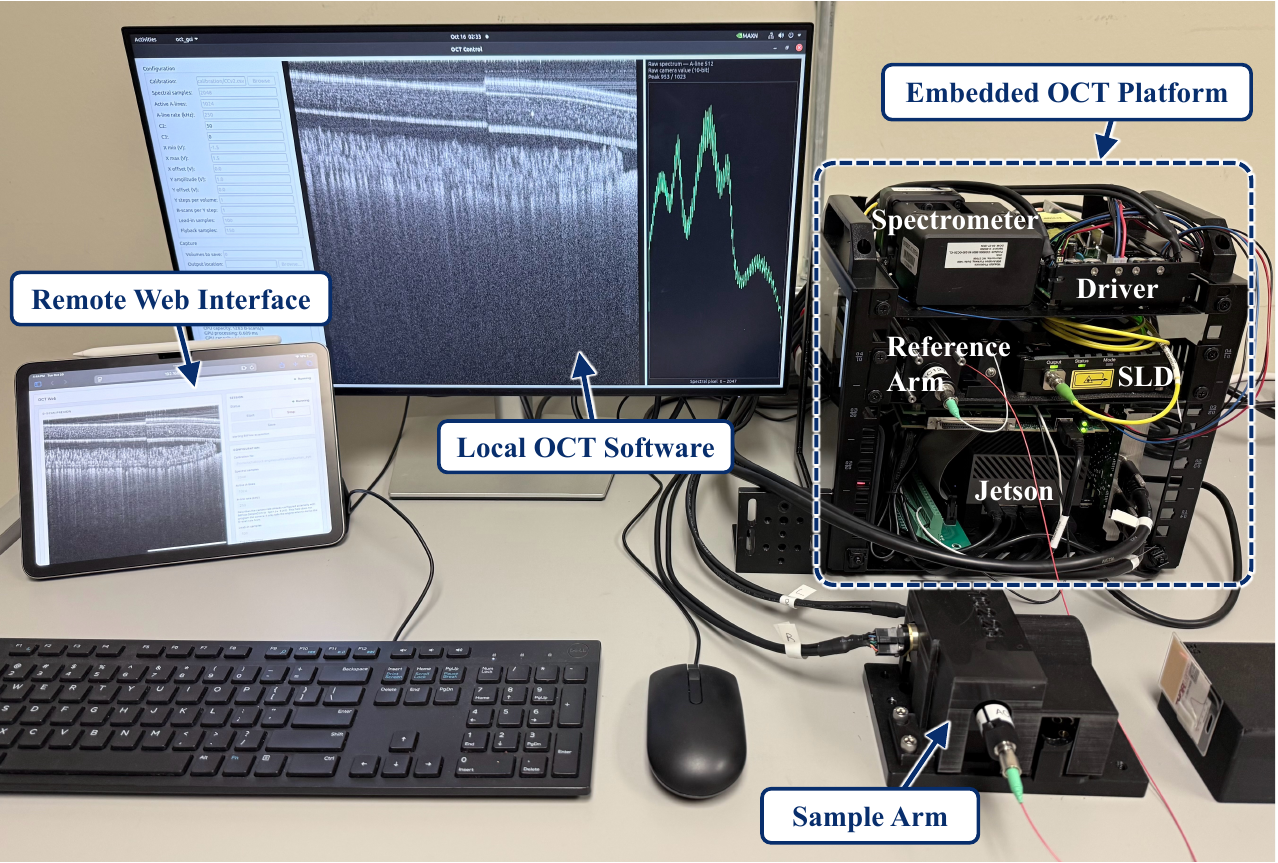}
    \caption{Photograph of the fully integrated embedded OCT imaging platform. The OCT system can operate entirely through a tablet interface, which provides system control, data acquisition, and real-time image visualization without the need for a dedicated monitor, keyboard, or mouse. The electronics footprint shown here does not represent the minimum achievable system size. Additional size reductions are feasible through integration of the DAQ, driver electronics, and control functions into compact microcontroller-based designs.}
    \label{fig:system}
\end{figure}

\begin{figure*}[!t]
    \centering
    \includegraphics[width=0.8\textwidth]{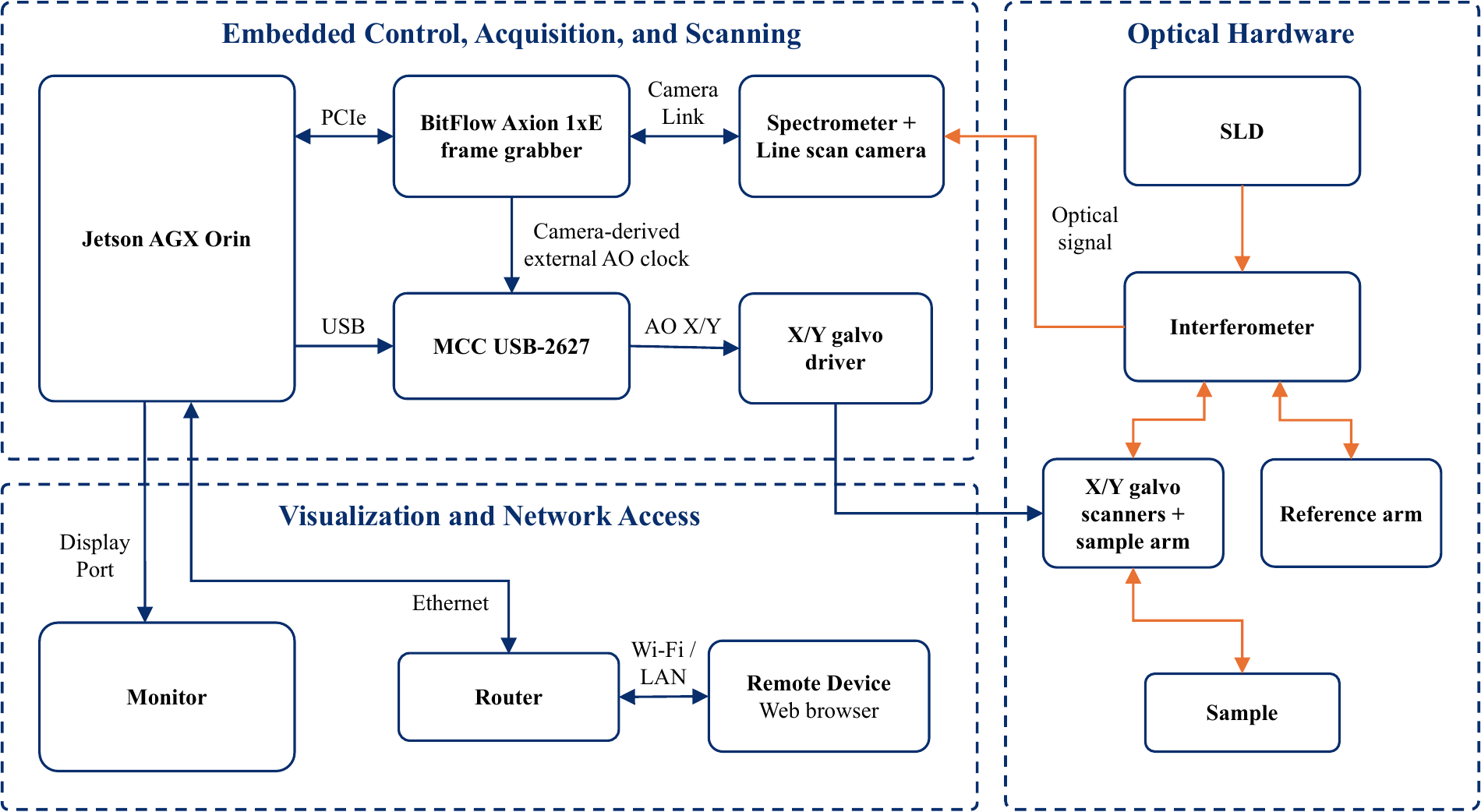}
    \caption{Hardware architecture of the embedded SD-OCT system. The NVIDIA Jetson AGX Orin directly interfaces with the Camera Link spectrometer through a PCIe frame grabber and controls the X/Y galvanometer scanners through a USB-connected DAQ synchronized to camera-derived timing signals. The optical subsystem consists of a superluminescent laser diode (SLD), a fiber-based Michelson interferometer, and sample and reference arms. Local and network-connected interfaces provide image visualization and system control.}
    \label{fig:hardware}
\end{figure*}

\begin{figure*}[!t]
    \centering
    \includegraphics[width=0.8\textwidth]{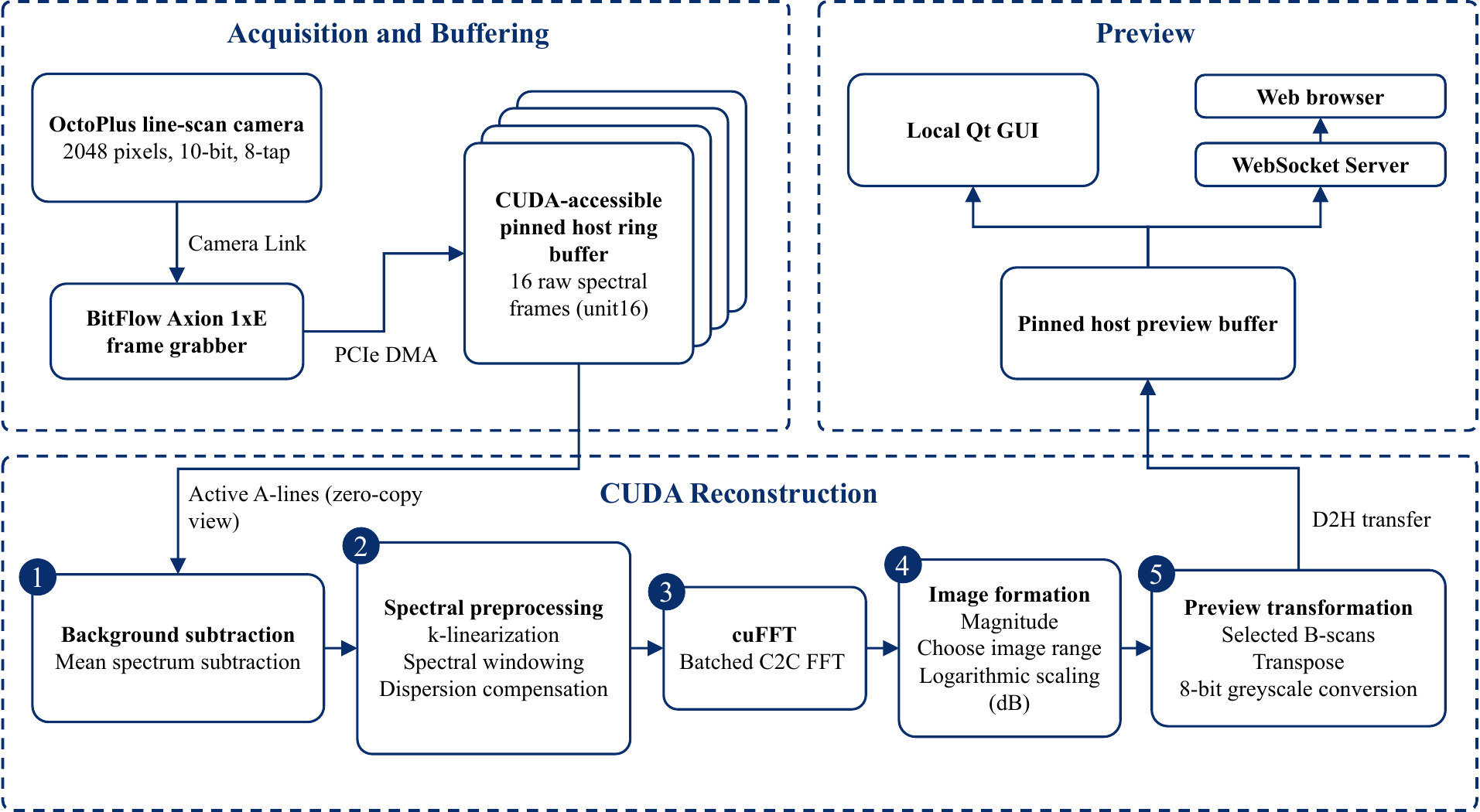}
    \caption{Embedded OCT acquisition, reconstruction, and visualization pipeline. Raw spectra are transferred by PCIe DMA into a CUDA-accessible pinned host ring buffer. Every active A-line undergoes four full-resolution GPU stages: (1) background subtraction, (2) spectral preprocessing, (3) cuFFT, and (4) logarithmic image formation. Selected B-scans additionally undergo preview transformation and host download for local or browser-based visualization.}
    \label{fig:pipeline}
\end{figure*}

Two-axis galvanometer scanning (ScannerMax, Sanford, FL, USA) was driven by two 16-bit, ±10-V analog outputs of a USB-2627 multifunction data acquisition device. Camera line-valid events were converted by the Axion timing sequencer into 1-µs pulses that externally clocked the USB-2627 analog-output scan. This hardware-timed arrangement updated both X/Y galvanometer control signals once per acquired A-line, eliminating dependence on software or USB timing. The optical hardware followed a conventional fiber-based Michelson interferometer comprising a 19 mm focal length fiber collimator, polarization controllers, a retroreflective mirror for a reference arm, and a sample arm consisting of a galvanometer scanner and a 64 mm focal length objective lens. The OCT system electronics were integrated into a compact 4U rack-mounted enclosure, reducing system footprint at 28 cm × 20 cm × 28 cm. The integrated OCT platform is shown in Fig.~\ref{fig:system}, while its hardware architecture and interconnections are illustrated in Fig.~\ref{fig:hardware}.

\subsection{OCT acquisition and reconstruction pipeline}
The acquisition and reconstruction software was implemented in C++17 and CUDA and executed natively on the Jetson AGX Orin running Ubuntu 20.04.6 LTS (Arm64), NVIDIA JetPack SDK 5.1.2 (Linux for Tegra R35.4.1), and CUDA Toolkit 11.4. At a 250-kHz A-scan rate, 2048 samples stored as 16-bit values corresponded to a nominal system-memory data rate of 1.024 GB/s. The BitFlow frame grabber transferred camera frames into a 16-entry circular ring of CUDA-accessible, page-locked host buffers. Taking advantage of the Jetson’s integrated CPU–GPU memory architecture, CUDA kernels accessed the active spectral data directly from these buffers, eliminating an explicit host-to-device copy for each B-scan. Lead-in and flyback A-lines were excluded through a non-owning view of the active region without an additional memory copy.

All device buffers, calibration, and the cuFFT steps were allocated during initialization. As summarized in Fig.~\ref{fig:pipeline}, the standard OCT reconstruction consisted of background subtraction, k-linearization, spectral windowing, numerical dispersion compensation, batched FFT, and logarithmic image formation. The full-resolution reconstructed image remained GPU-resident. Live B-scan visualization was decoupled from full-rate reconstruction. Selected B-scans were converted on the GPU to 8-bit grayscale at a target preview rate of 30 Hz and transferred to a reusable page-locked host buffer. A bounded latest-frame handoff prevented slow local or network displays from producing an accumulating preview queue. Frame identifiers were monitored throughout acquisition to detect discontinuities in the acquired sequence.

\subsection{Performance characterization}
To compare performance, we evaluated the Jetson and workstation platforms, whose specifications are summarized in Table~\ref{tab:platform_specs}. The Jetson was operated in MAXN mode with \texttt{jetson\_clocks} enabled during all performance measurements. The workstation used the Windows High Performance power plan, with the NVIDIA Control Panel power management mode set to "Prefer maximum performance." Two complementary measurement protocols were used: an equivalent visualization disabled platform comparison and a sustained GUI-enabled Jetson measurement.

For the matched comparison, the same C++/CUDA reconstruction implementation, processing parameters, timing definitions, and statistical aggregation were used on the Jetson and Windows workstation. Platform-specific code was limited to frame acquisition, scan control, and raw-data preparation. The workstation used a Teledyne DALSA Xtium2-CL MX4 frame grabber controlled through the Sapera LT SDK. To isolate the common acquisition-to-reconstruction path and prevent platform-specific display, network, and storage workloads from influencing the timing measurements, preview generation and download, rendering, network communication, and raw-data saving were disabled. Both systems used live camera input at an A-scan rate of 250-kHz with 2048 spectral samples per A-scan, 150 flyback samples, and either 512 or 1024 active A-scans per B-scan. Three independent 10-min runs were performed for each platform and B-scan size.

Wall power was measured at the AC input using the same TP-Link Kasa KP115 Wi-Fi smart plug for both platforms. Measurements included the computing platform and a line scan camera powered by PoCL (Power over Camera Link). Idle power was measured with the system configured and ready for acquisition, while active power was measured during continuous visualization disabled acquisition and reconstruction using 1024 active A-scans per B-scan. Each reported value was averaged over a 10-min interval. 

Processing time was measured for each B-scan after acquisition was complete. Three timing metrics were evaluated: data preparation time ($T_{\mathrm{prep}}$), reconstruction time ($T_{\mathrm{rec}}$), and total B-scan processing latency ($T_{\mathrm{proc}}$).

Data preparation time was measured from when a B-scan became available to the acquisition software ($t_{\mathrm{ready}}$) until its data were ready for GPU processing ($t_{\mathrm{in}}$). Reconstruction time was measured using host timestamps immediately before ($t_{\mathrm{rec},0}$) and after ($t_{\mathrm{rec},1}$) synchronized GPU reconstruction.
For B-scan $j$, these metrics were defined as
\begin{align}
T_{\mathrm{prep},j}
  &= t_{\mathrm{in},j}-t_{\mathrm{ready},j},
  \label{eq:staging_time}\\
T_{\mathrm{rec},j}
  &= t_{\mathrm{rec},1,j}-t_{\mathrm{rec},0,j},
  \label{eq:reconstruction_time}\\
T_{\mathrm{proc},j}
  &= t_{\mathrm{rec},1,j}-t_{\mathrm{ready},j}.
  \label{eq:processing_latency}
\end{align}
Total processing latency included data preparation, host scheduling between preparation and reconstruction, and reconstruction itself. It excluded camera acquisition and image display time.

On the workstation, data preparation included transferring the acquired B-scan from system memory to GPU memory. This upload time ($T_{\mathrm{up}}$) was measured separately but was already included in $T_{\mathrm{prep}}$. On the Jetson, explicit upload was unnecessary because the GPU directly accessed the acquisition buffers.

Frame availability was detected through the Sapera frame-completion callback on Windows and through the BitFlow acquisition thread on the Jetson. Therefore, the processing latencies reflect the software acquisition paths of the respective platforms rather than identical hardware-level timing boundaries.

GPU execution time was measured using CUDA events and defined as
\begin{equation}
T_{\mathrm{GPU},j}
=
T_{\mathrm{bg},j}
+
T_{\mathrm{pre},j}
+
T_{\mathrm{FFT},j}
+
T_{\mathrm{img},j},
\label{eq:gpu_time}
\end{equation}
where the terms represent background subtraction, spectral preprocessing, cuFFT, and logarithmic image formation, respectively. \(T_{\mathrm{GPU}}\) excluded preview conversion and download in the visualization disabled comparison. In contrast, \(T_{\mathrm{rec}}\) included kernel launches, GPU execution, synchronization, and host scheduling and should not be interpreted as CPU execution time.

GPU-equivalent processing rates reflect GPU execution time only, whereas B-scan processing rates also include acquisition data preparation, host scheduling, and reconstruction.

Timing statistics were collected over consecutive blocks of 10,000 B-scans. Results are reported as the mean $\pm$ sample standard deviation across the three runs. The 95th-percentile ($P_{95}$) and 99th-percentile ($P_{99}$) values were calculated for each complete block and then averaged. Final partial blocks were included in mean calculations and frame counts but excluded from percentile calculations.

Sustained GUI-enabled operation was evaluated independently from the GPU performance comparison. The Jetson performed one uninterrupted 800,000-B-scan run at 250-kHz using 1024 active A-scans per B-scan, corresponding to 62.6 min of acquisition including flyback samples. The local GUI and 30-Hz preview remained active. Frame identifiers were monitored for sequence discontinuities, and clock frequency, utilization, temperature, and power-rail measurements were periodically obtained using \texttt{tegrastats}.

Axial resolution was characterized using a mirror and its full width at half maximum (FWHM) of point-spread function (PSF). Lateral resolution was characterized using a 1951 USAF resolution target and line spread function FWHM.

\begin{table*}[!t]
\caption{Specifications of the embedded and workstation computing platforms used for performance measurements.}
\label{tab:platform_specs}
\centering
\begin{tabular*}{1.0\textwidth}{@{\hspace{8pt}\extracolsep{\fill}}lll@{\hspace{8pt}}}
\toprule[1.2pt]
Specification & Jetson AGX Orin & Workstation \\
\midrule[0.4pt]
CPU & 12-core Arm Cortex-A78AE & Intel Core i9-13900K \\
CPU cores / threads & 12 / 12 & 24 / 32 \\
GPU & NVIDIA Jetson AGX Orin Ampere GPU & NVIDIA RTX A4500 \\
CUDA cores & 2048 & 7168 \\
System memory & 64 GB unified LPDDR5 & 128 GB System RAM \\
GPU memory & Shared unified memory & 20 GB GDDR6 \\
Cost & \$2,000-\$3,500 USD & \$8,000-\$10,000 USD \\
\bottomrule[1.2pt]
\end{tabular*}
\end{table*}

\subsection{Web-based remote operation}
To enable fully embedded OCT operation without requiring a dedicated monitor, keyboard, or mouse, an asynchronous HTTP/WebSocket server was implemented using Boost.Asio/Beast and hosted directly on the Jetson. The server provided a browser-based interface over the local network for configuring scan parameters, controlling acquisition and raw-data capture, and monitoring system status and processing diagnostics. Control and diagnostic information were exchanged as JSON text messages, while B-scan previews were transmitted as binary WebSocket messages. Client-supplied acquisition and processing parameters were validated by the OCT engine before acquisition began.

Selected B-scans were converted on the GPU into 8-bit grayscale previews at a configurable target rate of 30 Hz. The resulting pixel images were transmitted as uncompressed binary data containing image dimensions and row stride. Acquisition and reconstruction executed on a dedicated worker thread, while network communication was handled by a separate asynchronous event loop. To prevent network backpressure from affecting acquisition, the outgoing preview queue was limited to one pending frame, with additional previews discarded rather than accumulated. Disconnection of the active browser client initiated an orderly acquisition stop and hardware shutdown.

\subsection{Sample imaging}
An infrared laser card (Thorlabs, NJ, USA) was imaged as a sample. Human subject imaging was performed under a protocol approved by the University at Buffalo Institutional Review Board (IRB). The tympanic membrane (eardrum) and fingertip of a healthy adult volunteer with no evidence of otologic disease were imaged using the embedded OCT platform, employing the ear imaging configuration described in previous work \cite{RN37}. Volumetric datasets consisting of 1024 × 1024 A-scans per volume over 6 × 6 mm field of view were acquired.

\section{Results}
\subsection{Matched reconstruction performance}
Across three 10-min measurements per platform and B-scan size, the Jetson and workstation achieved 378 B-scans/s with 512 active A-scans and 213 B-scans/s with 1024 active A-scans. No frame loss occurred in any matched run.

The GPU-stage measurements are summarized in Table~\ref{tab:gpu_performance}. The workstation reduced mean GPU execution time by 53.5\% and 57.2\% for the 512- and 1024-A-scan configurations, respectively, consistent with the higher computational capability of the RTX A4500.

\begin{table*}[!t]
\caption{GPU reconstruction performance under visualization disabled operation.}
\label{tab:gpu_performance}
\centering
\begin{tabular*}{\textwidth}{@{\hspace{8pt}\extracolsep{\fill}}lcccc@{\hspace{8pt}}}
\toprule[1.2pt]
& \multicolumn{2}{c}{512 active A-scans} &
  \multicolumn{2}{c}{1024 active A-scans} \\
\cmidrule(lr){2-3}\cmidrule(lr){4-5}
Processing stage &
Jetson & Workstation & Jetson & Workstation \\
\midrule[0.4pt]
\stagenum{1} Background subtraction (ms)
& $0.07920\pm0.00020$ & $0.06160\pm0.00040$
& $0.14080\pm0.00020$ & $0.09430\pm0.00200$ \\
\addlinespace[1pt]
\stagenum{2} Spectral preprocessing (ms)
& $0.07700\pm0.00080$ & $0.02792\pm0.00003$
& $0.14810\pm0.00020$ & $0.04971\pm0.00004$ \\
\addlinespace[1pt]
\stagenum{3} cuFFT (ms)
& $0.09540\pm0.00040$ & $0.03315\pm0.00005$
& $0.18310\pm0.00030$ & $0.06367\pm0.00011$ \\
\addlinespace[1pt]
\stagenum{4} Image formation (ms)
& $0.04453\pm0.00009$ & $0.01503\pm0.00003$
& $0.07996\pm0.00013$ & $0.02865\pm0.00002$ \\
\midrule[0.4pt]
Total $T_{\mathrm{GPU}}$ (ms)
& $0.29620\pm0.00100$ & $0.13770\pm0.00040$
& $0.55200\pm0.00050$ & $0.23630\pm0.00220$ \\
GPU-equivalent B-scan processing rate (kHz)
& 3.376 & 7.260 & 1.812 & 4.232 \\
GPU-equivalent A-scan processing rate (MHz)
& 1.729 & 3.717 & 1.855 & 4.333 \\
\bottomrule[1.2pt]
\end{tabular*}
\end{table*}

\subsection{B-scan processing latency}
The acquisition rates and processing latencies measured after each B-scan became available to the acquisition software are summarized in Table~\ref{tab:platform_timing}. The Jetson eliminated explicit host-to-device upload, whereas the workstation incurred mean upload latencies of 0.1182 and 0.2108 ms for 512 and 1024 active A-scans, respectively. Despite this overhead, the workstation achieved lower mean reconstruction and processing latencies, as well as lower mean $P_{99}$ processing latency. All reported mean and $P_{99}$ processing latencies remained below the corresponding B-scan acquisition intervals at 250-kHz. No dropped B-scans were reported by the acquisition software.

\begin{table*}[!t]
\caption{Acquisition, GPU reconstruction, and B-scan processing performance under visualization disabled operation.}
\label{tab:platform_timing}
\centering
\begin{tabular*}{\textwidth}{@{\hspace{8pt}\extracolsep{\fill}}lcccc@{\hspace{8pt}}}
\toprule[1.2pt]
& \multicolumn{2}{c}{512 active A-scans} &
  \multicolumn{2}{c}{1024 active A-scans} \\
\cmidrule(lr){2-3}\cmidrule(lr){4-5}
Metric & Jetson & Workstation & Jetson & Workstation \\
\midrule[0.4pt]
Acquired B-scans
& 679806 & 679803 & 383327 & 383316 \\
Acquired B-scan rate (Hz)
& 377.646 & 377.649 & 212.948 & 212.950 \\
B-scan acquisition interval (ms)
& 2.648 & 2.648 & 4.696 & 4.696 \\
\midrule[0.4pt]
Mean $T_{\mathrm{prep}}$ (ms)
& $0.0015\pm0.0002$ & $0.1504\pm0.0075$
& $0.0015\pm0.0001$ & $0.2339\pm0.0032$ \\
Mean $T_{\mathrm{up}}$ (subset of $T_{\mathrm{prep}}$) (ms)
& n.a. & $0.1182\pm0.0016$
& n.a. & $0.2108\pm0.0018$ \\
Mean $T_{\mathrm{rec}}$ (ms)
& $0.6317\pm0.0013$ & $0.1562\pm0.0020$
& $0.8871\pm0.0006$ & $0.2559\pm0.0025$ \\
Mean $T_{\mathrm{proc}}$ (ms)
& $0.6337\pm0.0012$ & $0.3088\pm0.0062$
& $0.8890\pm0.0007$ & $0.4922\pm0.0010$ \\
\midrule[0.4pt]
$P_{95}$ of $T_{\mathrm{proc}}$ (ms) [\% above mean]
& 0.6398 [0.96\%] & 0.3479 [12.7\%] & 0.8953 [0.71\%] & 0.5598 [13.7\%] \\
$P_{99}$ of $T_{\mathrm{proc}}$ (ms) [\% above mean]
& 0.6448 [1.75\%] & 0.4253 [37.7\%] & 0.9000 [1.23\%] & 0.6633 [34.8\%] \\
\midrule[0.4pt]
Equivalent B-scan processing rate (Hz)
& 1578.1 & 3237.9 & 1124.9 & 2031.6 \\
Equivalent A-scan processing rate (kHz)
& 808.0 & 1657.8 & 1151.9 & 2080.3 \\

\bottomrule[1.2pt]
\end{tabular*}
\end{table*}

\subsection{Wall-power and energy efficiency}
Wall-power and energy-efficiency measurements for the matched 1024-A-scan workload are summarized in Table~\ref{tab:power_efficiency}. The Jetson exhibited approximately 57\% lower indicated active wall power and 2.3-fold greater indicated wall-energy efficiency than the workstation.

\begin{table}[!t]
\caption{Wall power and energy efficiency during operation}
\label{tab:power_efficiency}
\centering
\begin{tabular}{lcc}
\toprule[1.2pt]
Metric & Jetson & Workstation \\
\midrule[0.4pt]
Idle wall power (W) & 56 & 120 \\
Active wall power (W) & 67 & 155 \\
B-scan rate (Hz) & 213 & 213 \\
$E_{\mathrm{B}}$ (mJ/B-scan) & 315 & 728 \\
$\eta_{\mathrm{B}}$ (B-scans/J) & 3.18 & 1.37 \\
\bottomrule[1.2pt]
\end{tabular}
\end{table}

\subsection{Sustained GUI-enabled performance}
During 62.6 min of continuous GUI-enabled acquisition, the embedded system processed 800,000 B-scans, corresponding to 819.2 million active A-scans, without a gap in the acquired frame sequence. Across the sampled 10,000-B-scan blocks, mean GPU processing time, including amortized preview conversion, ranged from 0.571 to 0.629 ms. Mean host-observed GUI-enabled processing wall time, including amortized preview conversion and preview download, ranged from 0.763 to 0.947 ms, with no sustained increase in processing latency. The maximum recorded processing latency was 5.328 ms, exceeding the B-scan acquisition interval of 4.696 ms; however, the buffered acquisition pipeline accommodated this isolated delay without frame loss.

Periodic system measurements for the Jetson showed that CPU and GPU temperatures stabilized near 58~$^\circ$C and 53~$^\circ$C, respectively, without an observed sustained reduction in clock frequencies. GPU utilization ranged from approximately 19\% to 44\% across sampled readings, while most CPU cores remained below 15\% utilization. Sampled internal power-rail measurements indicated approximately 20 W of Jetson module power consumption. Because these internal measurements do not share the same measurement boundary as workstation wall power, they were used only to monitor the Jetson operating state and were not used in the cross-platform energy-efficiency calculations.

\begin{figure}[!t]
    \centering
    \includegraphics[width=\columnwidth]{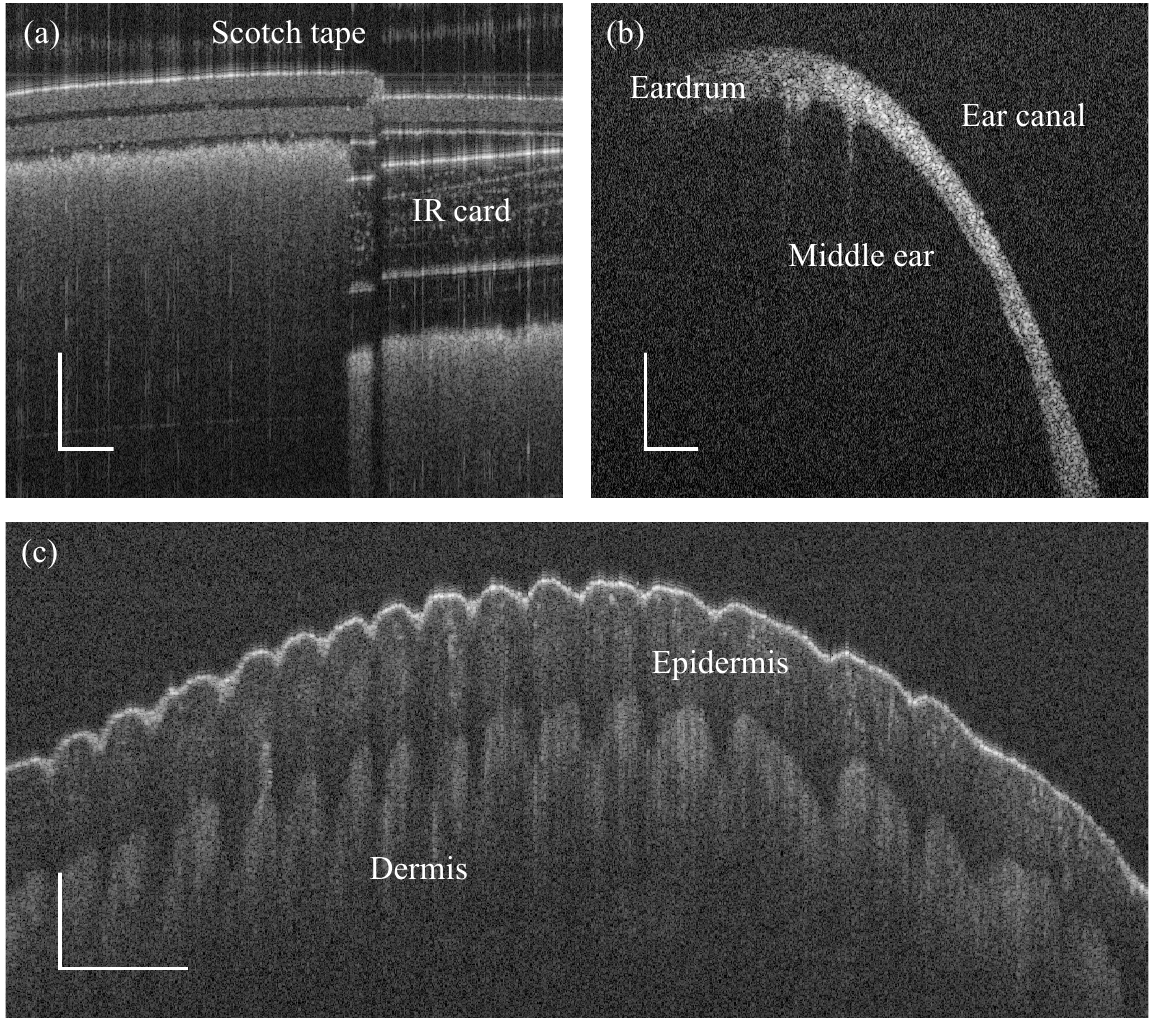}
    \caption{Representative OCT B-scans acquired using the embedded 250-kHz SD-OCT system. (a) Multiple layers of Scotch tape on an infrared laser card, (b) human tympanic membrane (eardrum), and (c) fingertip. Scale bars represent 200~$\mu$m.}
    \label{fig:imaging_demo}
\end{figure}

\subsection{Imaging and remote operation demonstration}
The web interface enabled remote configuration, acquisition control, raw-data capture, and live B-scan visualization from a tablet connected to the local network (Fig.~\ref{fig:system}). Acquisition and GPU reconstruction continued while processed B-scans were transmitted to the browser for visualization.

Representative B-scans of an infrared laser card, human tympanic membrane, and fingertip are shown in Fig.~\ref{fig:imaging_demo}. Mirror measurements yielded an axial resolution of 3.41 µm in air. The lateral line spread function FWHM was 12.3 µm, with a USAF target resolution of 8.77 µm.

\section{Discussion}
This work demonstrates that 250-kHz OCT acquisition, reconstruction, visualization, and hardware control can be integrated within an embedded Jetson platform without a workstation. The performance comparison showed that the workstation provided lower mean processing latency, whereas the Jetson eliminated explicit host-to-device upload and exhibited substantially lower processing latency variability, with $P_{95}$ and $P_{99}$ values remaining within 2\% of the mean processing latency, compared with deviations of up to 38\% on the workstation. The wall-power measurements also demonstrate an important system-level advantage of the embedded architecture. Using the same scan protocols of 1024-A-scan per B-scan, the power measurements corresponded to approximately 57\% lower active wall power and approximately 2.3-fold greater active wall-energy efficiency for the Jetson. In particular, both systems were evaluated in maximum performance configurations despite processing rates considerably exceeding acquisition requirements. Consequently, further reductions in power consumption may be possible without compromising real-time operation.

Compact OCT implementations have historically involved tradeoffs in acquisition speed due to constraints in detector speeds or computational throughput. For example, Kim et al. \cite{RN9} demonstrated a mini-PC-based OCT system operating at an A-scan rate of 8.8-kHz, while Dsouza et al. \cite{RN8} demonstrated a laptop-based OCT system operating at 10-kHz. Recently, Wang et al. \cite{RN15} demonstrated 80-kHz acquisition using a USB3-connected camera and Jetson Orin Nano, and suggested PCIe-based Camera Link acquisition as a potential approach for supporting higher-speed spectrometers. In the present work, a PCIe Camera Link frame grabber was integrated directly into an embedded Jetson platform, enabling operation with a line-scan camera providing a 3.1-fold higher acquisition rate and twice the spectral sampling density. These results demonstrate that acquisition bandwidth traditionally associated with workstation-based OCT systems can also be achieved within an embedded computing architecture.

The capability to support high acquisition speed provides important advantages in translational and clinical OCT imaging, provided that sufficient sensitivity, imaging range, and system performance are maintained. This is particularly relevant as modern commercial OCT instruments increasingly operate at acquisition rates in the 100-400-kHz range to support ultrawide volumetric imaging, reduced patient and operator motion, and the increased number of repeated B-scans for OCTA and other functional OCT methods \cite{RN28,RN29}. Although the demonstrated imaging platform was limited by the maximum line scan camera rate of 250-kHz, the measured host-side processing performance Table~\ref{tab:platform_timing} suggests that the embedded architecture is not the main bottleneck. Average CPU and GPU utilization also remained relatively low during 250-kHz operation, indicating that computational resources were not fully utilized. Based on the mean processing latency, the Jetson provided equivalent active A-scan processing rates of 808.0 kHz and 1151.9 kHz
for the 512- and 1024-A-scan configurations, respectively, indicating potential use for swept source OCT (SS-OCT) systems with high-speed digitizers.

Beyond acquisition and processing performance, embedded OCT platforms offer practical advantages including reduced physical footprint, lower wall-power demand, direct GPU processing that minimizes data-transfer overhead, and more consistent processing latency. The integration of computing, graphics, memory, and I/O into a single module also provides a lower price point than a workstation requiring separate high-performance components, including a CPU, discrete GPU, system memory, power supply, cooling system, and chassis. Operation through remote network access without a dedicated monitor, keyboard, or mouse may further facilitate deployment in portable, handheld, or space-constrained OCT systems. Furthermore, the heterogeneous CPU--GPU architecture of modern embedded systems is ideal to support parallel execution of computationally intensive imaging pipelines, potentially enabling multimodal acquisition, processing, and analysis within a single platform.

Several limitations should be acknowledged. First, experimental validation was limited to a maximum acquisition rate of 250-kHz by the line-scan camera, preventing direct evaluation at higher imaging speeds. Equivalent throughput values therefore describe computational performance rather than a validated acquisition limit. Demonstration at higher A-scan rates will require implementation and testing in SS-OCT systems. Second, although the same reconstruction pipeline, timing definitions, and statistical aggregation were used on both computing platforms, their acquisition paths remained platform specific. For example, the Jetson used direct CUDA access to BitFlow acquisition buffers, whereas the workstation used Sapera acquisition followed by an explicit host-to-device copy. Differences in frame-grabber hardware, memory architecture, operating system, and device drivers therefore remained part of the measured system-level processing latency.

Third, wall power was measured during one matched workload using a consumer-grade KP115 smart plug. The measurements quantify complete-system power within the selected boundary but do not isolate the GPU, CPU, PoCL for camera, or other components. Additional measurements using a calibrated power analyzer and scan protocols would be required for a complete uncertainty-qualified power--performance characterization.

Future work should focus on deployment in higher speed and multimodal OCT systems, where the computational demands of acquisition, reconstruction, visualization, and AI inference are more stringent. Embedded GPU platforms may offer a practical framework for integrating these functions within a single low-power system while preserving real-time operation. In addition, network-connected OCT systems offer opportunities for remote imaging and specialist-guided examinations. Together, also with advances in photonic integrated circuit-based OCT technologies \cite{RN38,RN39}, these approaches could facilitate the development of small, potentially battery-powered OCT instruments while retaining the computational capabilities required for high-performance OCT.

\section{Conclusion}
Embedded GPU computing supported real-time SD-OCT acquisition and processing at an A-scan rate of 250-kHz while substantially reducing power consumption and enabling a smaller, lower-cost computing platform compared with a conventional workstation. These findings support the feasibility of compact, high-performance OCT systems, including handheld OCT probes that eliminate the need for a bulky imaging cart and may improve clinical workflow. The demonstrated computational capacity also suggests the feasibility of higher-speed OCT systems and multimodal imaging on embedded platforms.

\section{Acknowledgments}
This research was supported by the National Center for Advancing Translational Sciences of the National Institutes of Health under award number K12TR005295 to the University at Buffalo (UB). The content is solely the responsibility of the authors and does not necessarily represent the official views of the NIH. This research was also supported in part by UB School of Engineering and Applied Sciences (SEAS) Summer Undergraduate Research Experience (SURE) Program and start-up research funds.

\bibliographystyle{IEEEtran}
\bibliography{references}


 





\end{document}